\documentclass[journal]{IEEEtran}
\usepackage{amsmath,graphicx,multirow,booktabs,amssymb,subfigure,cite,bm}

\begin{document}
%
\title{Gated Recurrent Fusion with Joint Training Framework for Robust End-to-End Speech Recognition}
%
%
%

\author{Cunhang~Fan,~\IEEEmembership{Student Member,~IEEE,}
        Jiangyan~Yi,~\IEEEmembership{Member,~IEEE,}
        Jianhua~Tao,~\IEEEmembership{Senior Member,~IEEE,}
        Zhengkun~Tian,~\IEEEmembership{Student Member,~IEEE,}
        Bin~Liu,~\IEEEmembership{Member,~IEEE,}
        and~Zhengqi~Wen,~\IEEEmembership{Member,~IEEE,}
\thanks{This work is supported by the National Key Research \& Development Plan of China (No.2017YFC0820602) and the National Natural Science Foundation of China (NSFC) (No.61425017, No.61831022, No.61773379, No.61771472), and Inria-CAS Joint Research Project (No.173211KYSB20170061).\emph{(Corresponding authors: Jiangyan Yi and Jianhua Tao.)}}
\thanks{The authors are with the National Laboratory of Pattern Recognition, Institute of Automation, Chinese Academy of Sciences, Beijing 100190, China. Cunhang Fan, Jianhua Tao and Zhengkun Tian are also with the School of Artificial Intelligence, University of Chinese Academy of Sciences, Beijing 100190, China. Jianhua Tao is also with the CAS Center for Excellence in Brain Science and Intelligence Technology, Beijing 100190, China. (e-mail:cunhang.fan@nlpr.ia.ac.cn, jiangyan.yi@nlpr.ia.ac.cn, jhtao@nlpr.ia.ac.cn, zhengkun.tian@nlpr.ia.ac.cn, liubin@nlpr.ia.ac.cn, zqwen@nlpr.ia.ac.cn)}
}

\markboth{Journal of \LaTeX\ Class Files,~Vol.~14, No.~8, August~2015}
{Shell \MakeLowercase{\textit{et al.}}: Bare Demo of IEEEtran.cls for IEEE Journals}

%



\maketitle

\begin{abstract}
The joint training framework for speech enhancement and recognition methods have obtained quite good performances for robust end-to-end automatic speech recognition (ASR). However, these methods only utilize the enhanced feature as the input of the speech recognition component, which are affected by the speech distortion problem. In order to address this problem, this paper proposes a gated recurrent fusion (GRF) method with joint training framework for robust end-to-end ASR. The GRF algorithm is used to dynamically combine the noisy and enhanced features. Therefore, the GRF can not only remove the noise signals from the enhanced features, but also learn the raw fine structures from the noisy features so that it can alleviate the speech distortion. The proposed method consists of speech enhancement, GRF and speech recognition. Firstly, the mask based speech enhancement network is applied to enhance the input speech. Secondly, the GRF is applied to address the speech distortion problem. Thirdly, to improve the performance of ASR, the state-of-the-art speech transformer algorithm is used as the speech recognition component. Finally, the joint training framework is utilized to optimize these three components, simultaneously. Our experiments are conducted on an open-source Mandarin speech corpus called AISHELL-1. Experimental results show that the proposed method achieves the relative character error rate (CER) reduction of 10.04\% over the conventional joint enhancement and transformer method only using the enhanced features. Especially for the low signal-to-noise ratio (0 dB), our proposed method can achieves better performances with 12.67\% CER reduction, which suggests the potential of our proposed method.
\end{abstract}


%
\IEEEpeerreviewmaketitle

\section{Introduction}
%
%
%
%

\IEEEPARstart{T}{here} are many methods for automatic speech recognition (ASR) systems, such as GMM-HMM and deep neural network (DNN) based acoustic models \cite{hinton2012deep,xiong2018microsoft,saon2017english,li2017acoustic}. Recently, end-to-end speech recognition methods \cite{graves2014towards,chorowski2015attention,dong2018speech,povey2018time,salazar2019self,karita2019improving,tian2019self} have made significantly breakthroughs. Although these ASR methods have made a lot of progresses on clean speech signals, the performance could be dramatically degraded in the noisy and reverberation environments. In realistic environments, recorded speech signals are always interfered by various background noises and reverberations. Therefore, improving the robustness of ASR is very important. This paper focuses on boosting the noise robustness of end-to-end speech recognition.


In order to boost the noise robustness of ASR, there are three mainstream methods. The first mainstream method is adding the speech enhancement component at the front-end of ASR. Speech enhancement methods include spectral subtraction \cite{boll1979suppression}, Wiener filtering \cite{scalart1996speech} and deep neural network (DNN) based speech enhancement methods \cite{vincent2008extracting,narayanan2013ideal,wang2014training,pascual2017segan,fan2019noise,fujimoto2019one}. However, speech enhancement optimizes their models to estimate the target speech, which is different from the speech recognition part. Therefore, speech enhancement methods fail to optimize towards the final objective, which leads to a suboptimal solution \cite{seltzer2008bridging}. In addition, the enhanced speech by these speech enhancement methods usually generates over-smoothed speech, which is the reason of speech distortion after speech enhancement. The speech distortion can degrade the performance of ASR \cite{wang2016joint}. Therefore, the performance of this approach is highly dependent on the performance of the enhancement front-end \cite{han2015learning}.

The second mainstream method uses the multi-condition training (MCT) to boost the noise robustness of ASR. MCT uses different kinds of data (clean and noisy speech) to train the speech recognition model. However, the complexity and computing costs of MCT are increased. In addition, it gives unimpressive performance on the unmatched conditions \cite{li2014long} and the performance is also affected by the speech distortion \cite{seltzer2013investigation}. In order to alleviate speech distortion problem, the enhancement front-end enhances both training and test set first, and ASR model is trained with the enhanced data. It can improve the ASR performance in some degree, but it still highly depends on the performance of the enhancement front-end. Different from the MCT method, the SpecAugment \cite{park2019specaugment} directly applies the  data augmentation to the input features of neural networks (i.e., filter bank coefficients). The SpecAugment is used only during the training, which consists of three spectrogram deformations:  time warping,  time and frequency masking. Although the SpecAugment can improve the performance of end-to-end ASR, it needs to be improved on the noisy condition.

The third mainstream method is the joint training methods \cite{wang2016joint,liu2019jointly,liu2018boosting,chang2020end}. These methods apply the joint training framework to optimize the speech enhancement and recognition, simultaneously. The reason is that speech enhancement and speech recognition are not two independent tasks and they can clearly benefit from each other. In order to boost noise robustness of end-to-end ASR, in \cite{liu2019jointly}, authors propose a joint adversarial enhancement training method. They utilize the joint training framework to optimize the mask based enhancement network and attention based encoder-decoder speech recognition network. However, this method only uses the enhanced feature as the input of speech recognition, which is still affected by the speech distortion problem. In addition, in the noisy AISHELL-1 \cite{bu2017aishell} dataset, the character error rate (CER) of this method is still more than 50\%, which needs to be improved. As for the end-to-end speech recognition, speech transformer \cite{dong2018speech,povey2018time,salazar2019self,karita2019improving} models have shown impressive performance and acquired state-of-the-art results. Self-attention network \cite{vaswani2017attention} is one of the key components of speech transformer and it is more powerful to model long-term dependencies than recurrent neural networks (RNNs) based sequence to sequence models. Therefore, applying the joint training of enhancement and speech transformer can further improve the performance of robust end-to-end ASR.

In \cite{fujimoto2019one}, a one-pass robust speech recognition method is proposed. It combines the noisy and enhanced features by a gating mechanism. Although it can improve the robust of ASR, the enhancement and speech recognition are trained separately instead of the joint training algorithm. In addition, the simple gate mechanism can not make full use of the sequence information so that it can not fuse the noisy and enhanced features very well.

\begin{figure}[t]
	\centering
	\begin{minipage}[t]{0.45\textwidth}
		\includegraphics[width=\linewidth]{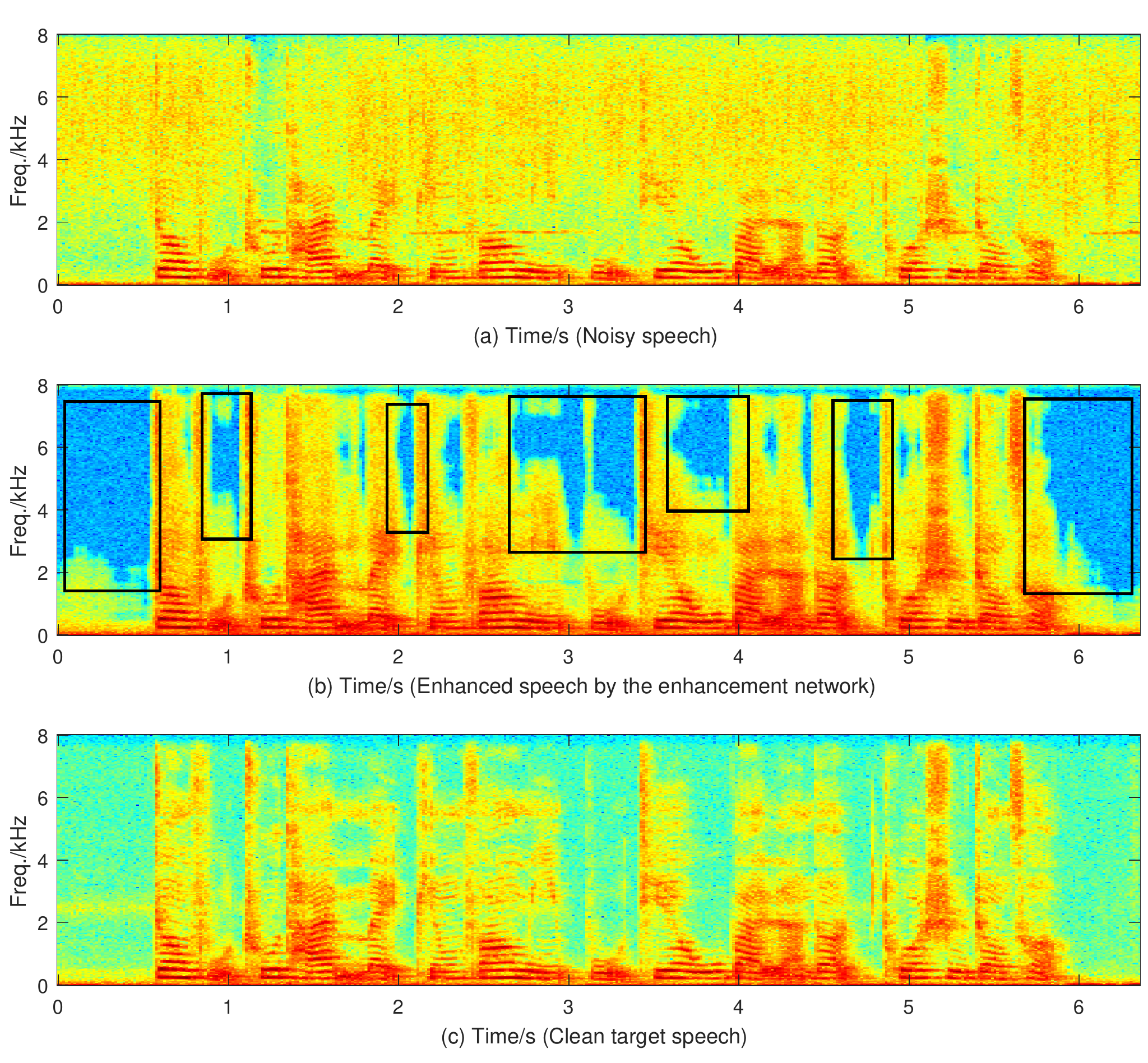}
	\end{minipage}
	\caption{The spectrogram example of a test speech sample. (a) The noisy speech. (b) The enhanced speech by the enhancement network, which is without the joint training. (c) The clean target speech.}
	\label{fig:yuputu}
\end{figure}

Fig.~\ref{fig:yuputu} illustrates the spectrogram example of a test speech sample. From Fig.~\ref{fig:yuputu} we can find that the spectrogram of the enhanced speech by the enhancement network has significant leaks (marked in Fig.~\ref{fig:yuputu} (b) by block boxes), which leads to the speech distortion. There are significant leaks in these black boxes. This is because that the noise is dominant in these T-F bins, which drowns the target speech. Therefore, the enhancement network deals with these T-F bins as the noise signals and removes most of the information. These leaks lose so much very important speech information, for example: formants. Although the enhancement network can remove noise signals in some degree, these leaks are unknown for the speech recognition system and lose so much speech information. These are the reasons why speech distortion damages the performance of speech recognition. 

In this paper, we propose a gated recurrent fusion (GRF) with joint training framework for robust end-to-end ASR. In order to address the speech distortion problem, motivated by \cite{liu20203d,fan2020gated}, the GRF is utilized to dynamically combine the noisy and enhanced features. Therefore, the GRF can offset these leaks from the noisy features. In addition, GRF can reduce the noise from the enhanced features. So the GRF aims to learn to adaptively select and fuse the relevant information from noisy and enhanced features by making full use of the gate and memory modules. The GRF can extract more appropriate and robust speech features. In addition, we apply the joint training algorithm to optimize the enhancement and speech recognition. The state-of-the-art end-to-end ASR method speech transformer with self-attention method is used as the speech recognition component. Specifically, the proposed joint training method includes three parts: speech enhancement, gated recurrent fusion and speech recognition. With the joint optimization of enhancement and recognition, the proposed model is expected to learn more robust representations suitable for the recognition task automatically. 

To summarize, the main contribution of this paper is two-fold. Firstly, to address the speech distortion problem, the gated recurrent fusion algorithm is utilized to dynamically fuse the noisy and enhanced features. Secondly, to the best of our knowledge, it is the first time to apply the speech transformer and single channel speech enhancement for the joint training framework. Our experiments are conducted on AISHELL-1 Mandarin dataset. Experimental results show that the proposed method achieves the relative CER reduction of 10.02\% over the conventional joint enhancement and transformer method using the enhanced features only. Especially for the low signal-to-noise ratios, our proposed method can achieve better performance.

The rest of this paper is organized as follows. Section \(\rm \uppercase\expandafter{\romannumeral2}\) presents the conventional joint training method for robust ASR. Section \(\rm \uppercase\expandafter{\romannumeral3}\) introduces our proposed joint training method with gated recurrent fusion algorithm. The experimental setup is stated in section \(\rm \uppercase\expandafter{\romannumeral4}\). Section \(\rm \uppercase\expandafter{\romannumeral5}\) shows experimental results. Section \(\rm \uppercase\expandafter{\romannumeral6}\) shows the discussions. Section \(\rm \uppercase\expandafter{\romannumeral7}\) draws conclusions.

\section{Conventional joint training method}
\label{sec:separation}

\begin{figure}[t]
	\centering
	\begin{minipage}[t]{0.3\textwidth}
		\includegraphics[width=\linewidth]{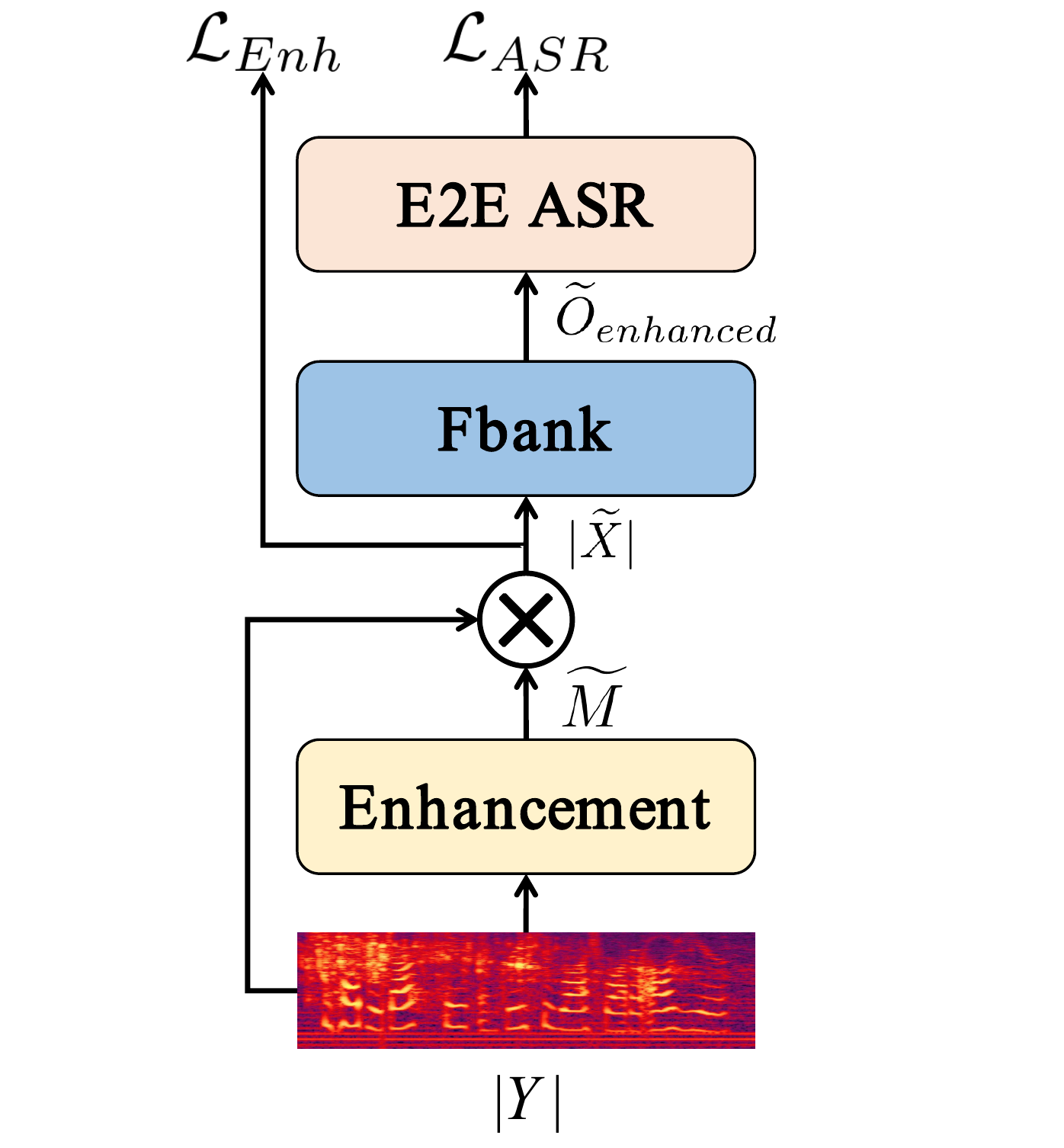}
	\end{minipage}
	\caption{The schematic diagram of conventional joint training method for robust end-to-end ASR.}
	\label{fig:baseline}
\end{figure}

Fig.~\ref{fig:baseline} illustrates the schematic diagram of conventional joint training method for robust end-to-end ASR. It usually consists of a enhancement network and an end-to-end ASR network. 

Speech enhancement aims to remove the noise signals and estimate the target clean speech from noisy input. The noisy speech can be represented as: 
	\begin{equation}
	y(t) = x(t)+n(t)
	\label{eq1}
	\end{equation}
where \(y(t)\), \(x(t)\) and \(n(t)\) denote the noisy, clean and noise signals, respectively. We apply the \(Y(t,f)\), \(X(t,f)\) and \(N(t,f)\) as the corresponding short-time Fourier transformation (STFT) of noisy, clean and noise signals, which still satisfy this relation:
\begin{equation}
Y(t,f) = X(t,f)+N(t,f)
\label{eq111}
\end{equation}
where \((t,f)\) denotes the index of time-frequency (T-F) bins. We drop the \((t,f)\) in the rest of this paper.

The noisy magnitude spectrum \(|Y|\) is used as the input feature. The conventional joint training robust end-to-end ASR system can be represented as follows:
\begin{equation}
|\widetilde{X}|=\rm Enhancement(|Y|)
\label{eq2}
\end{equation}
\begin{equation}
\widetilde{O}_{enhanced}=\rm Fbank(|\widetilde{X}|)
\label{eq3}
\end{equation}
\begin{equation}
\mathcal{L}_{ASR}=-lnP(S^*|\widetilde{O}_{enhanced})
\label{eq4}
\end{equation}
where, the \(\rm Enhancement(\cdot)\) denotes the speech enhancement function, which aims to estimate the clean target speech from the noisy input \(|Y|\). \(|\widetilde{X}|\) is the enhanced speech. The \(\rm Fbank(\cdot)\) denotes the function to extract Fbank features, which transforms the \(|\widetilde{X}|\) to \(\widetilde{O}_{enhanced}\). The \(\mathcal{L}_{ASR}\) is the loss function of end-to-end ASR. The posterior probabilities \(P(S^*|\widetilde{O}_{enhanced})\) of output labels \(S^*\) are estimated from the enhanced Fbank features \(\widetilde{O}_{enhanced}\).


As for the conventional joint training method, it includes two parts: the speech enhancement component and the speech recognition component. Firstly, the noisy and clean parallel data are used to train the speech enhancement model. Secondly, only the enhanced speech is utilized as the input feature of speech recognition model. Finally, the total loss of enhancement and speech recognition is applied to optimize the whole model. Therefore, the enhancement and ASR models can be jointly trained. However, this method only applies the enhanced feature as the input of speech recognition model, which is still affected by the speech distortion problem in some degree.

\section{Our proposed joint training method}

\begin{figure}[t]
	\centering
	\begin{minipage}[t]{0.3\textwidth}
		\includegraphics[width=\linewidth]{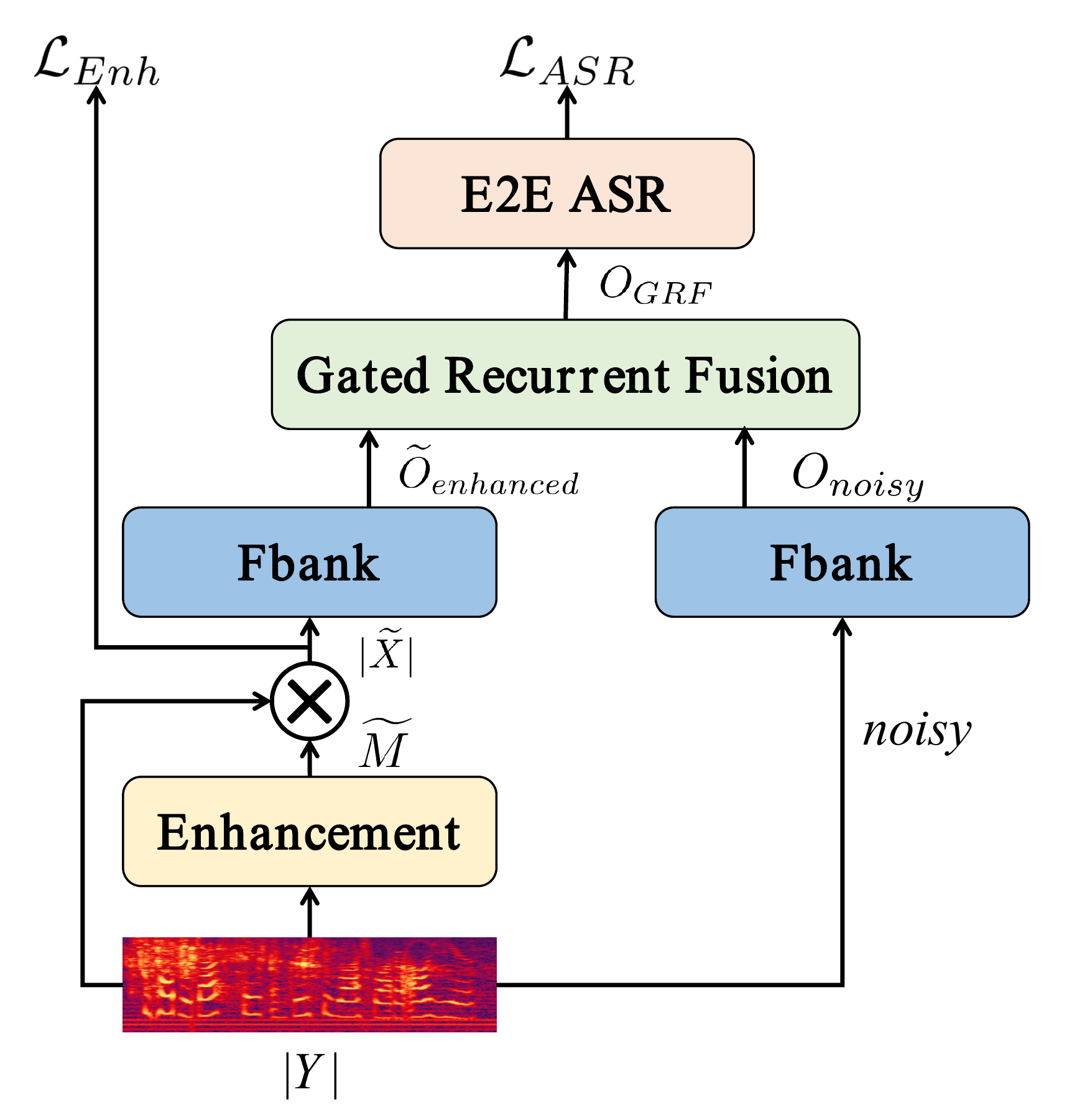}
	\end{minipage}
	\caption{The schematic diagram of our proposed joint training method for robust end-to-end ASR. To further alleviate speech distortion problem, our proposed method applies the gated recurrent fusion algorithm to combine the noisy and enhanced features.}
	\label{fig:proposed}
\end{figure}

Fig.~\ref{fig:proposed} illustrates the schematic diagram of our proposed gated recurrent fusion (GRF) with joint training framework for robust end-to-end ASR. It includes a mask-based speech enhancement network, a GRF network and an end-to-end ASR network. Firstly, the noisy magnitude spectrum \(|Y|\) is enhanced by the speech enhancement network. Secondly, the GRF is applied to extract GRF representations. Then these GRF representations are used as the input of end-to-end ASR network. Finally, we use a jointly compositional scheme to optimize the whole model, whose parameters are updated by stochastic gradient descent.

Specifically, we use the state-of-the-art end-to-end ASR method speech transformer as the end-to-end speech recognition component. In addition, to address the speech distortion problem, the GRF component is utilized to dynamically combine the noisy and enhanced features. Therefore, these GRF representations can learn the raw fine structures from the noisy features to alleviate the speech distortion. Meanwhile, they can also remove the noise signals from the enhanced features.

\begin{figure*}[t]
	\centering
	\subfigure[The overall schematic diagram of GRF]{
		\begin{minipage}[t]{0.4\linewidth}
			\centering
			\includegraphics[width=2.2in]{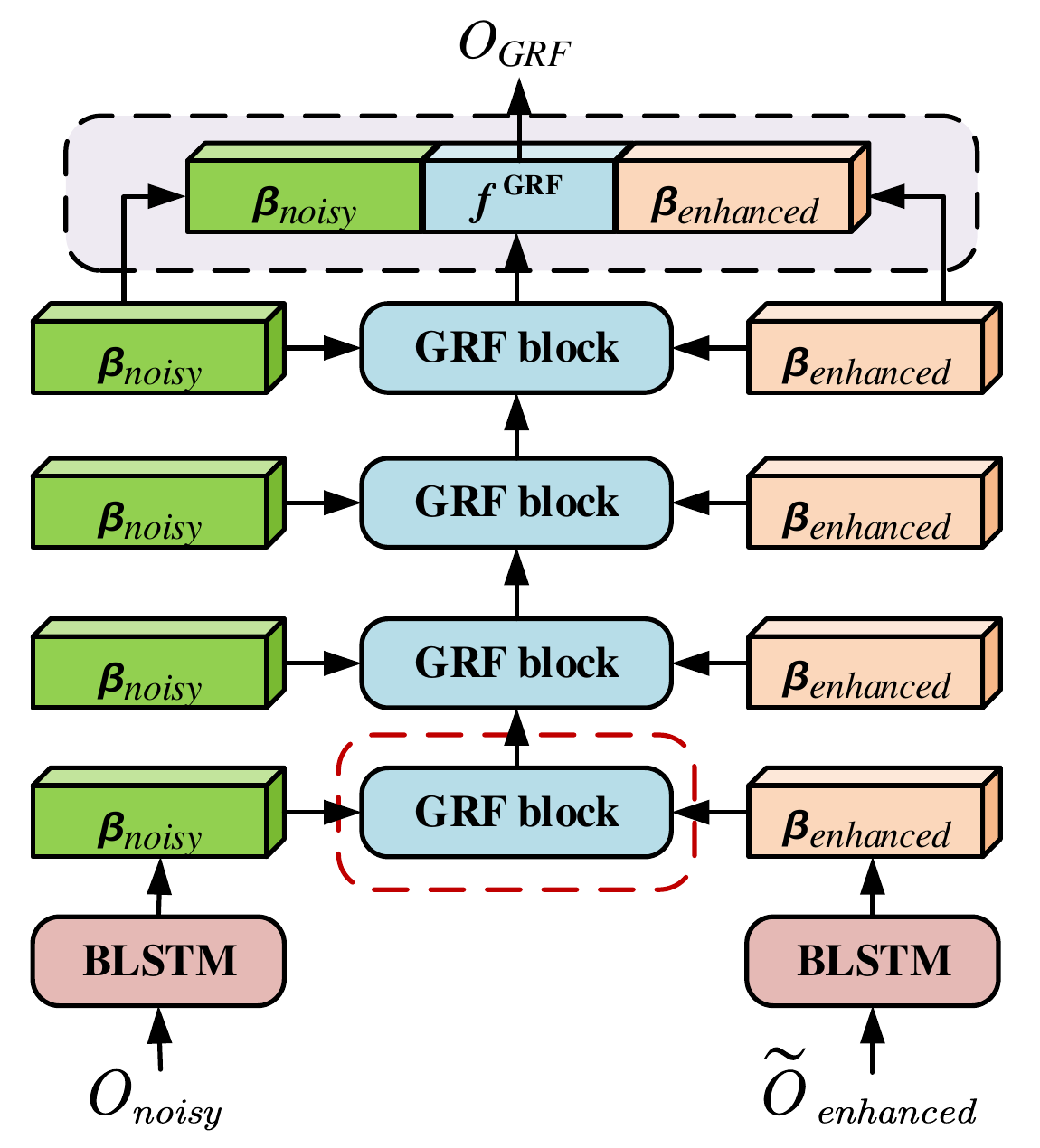}
		\end{minipage}%
		\label{fig:grf}
	}
	\subfigure[GRF block]{
		\begin{minipage}[t]{0.5\linewidth}
			\centering
			\includegraphics[width=3.8in]{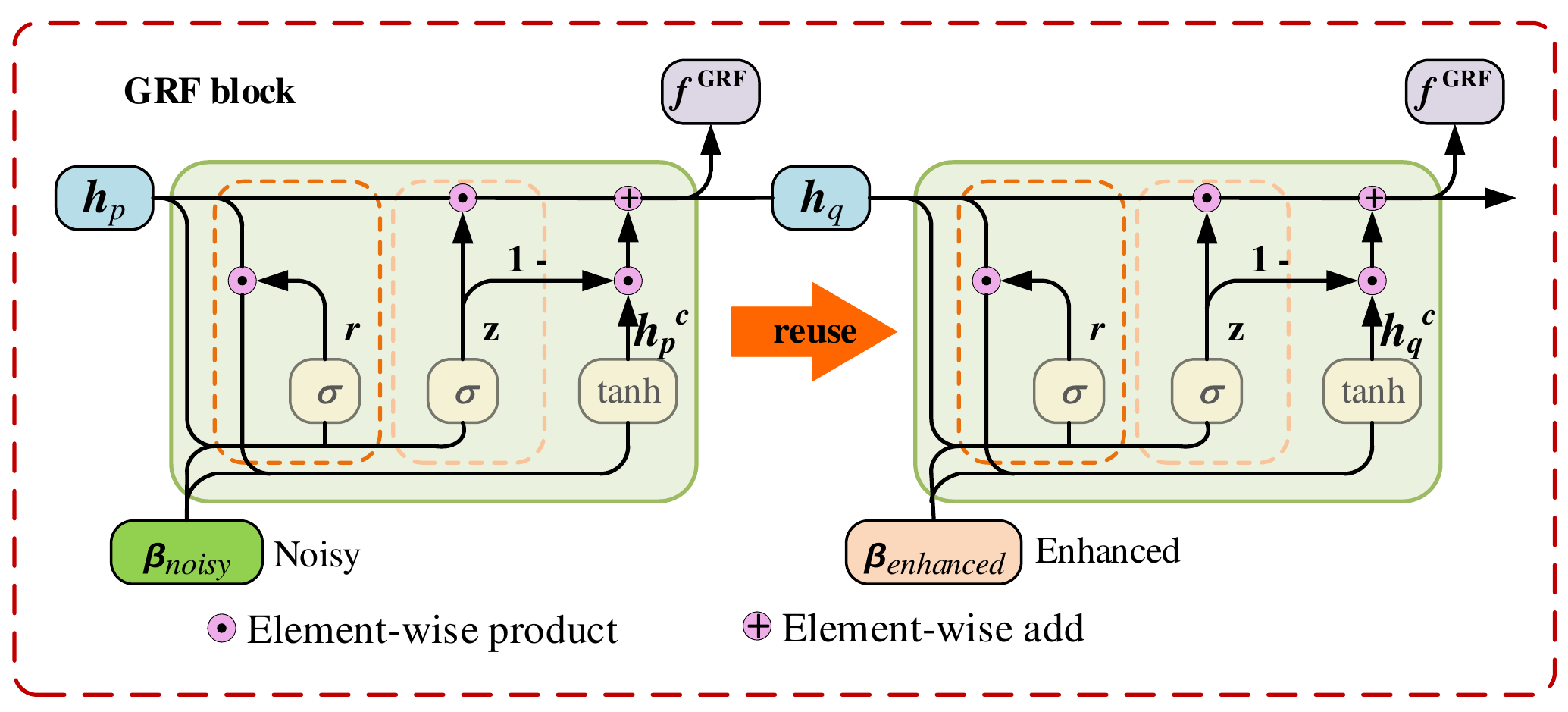}
		\end{minipage}%
		\label{fig:grf-block}
	}
	\centering
	\caption{(a) The overall schematic diagram of GRF, which fuses the noisy and enhanced features to address the speech distortion problem. (b) The schematic diagram of GRF block.}
	\label{fig:all}
\end{figure*}

\subsection{Mask-based enhancement network}
\label{sec:enhancement}

As for speech separation task, it is well known that mask-based speech separation can obtain a better result \cite{wang2018supervised,fan2019discriminative,fan2018Utterance,fan2020spatial,fan2020end,fan2020deep}. Similarly, we apply the mask \(M\) for speech enhancement. 

The main task of mask-based speech enhancement network is to estimate the target mask for each T-F bin. Then the estimated mask is applied to the noisy input to acquire the clean target speech. Because the Fbank features have nothing to do with the phase spectrums. In this paper, we utilize the ideal amplitude mask (IAM) \cite{kolbaek2017multitalker} rather than the phase sensitive mask (PSM) \cite{erdogan2015phase} for speech enhancement. The IAM is defined as follows:
\begin{equation}
M_{IAM}=\frac{|X|}{|Y|}
\label{eq511}
\end{equation}

For training, we use the spectrum approximation (SA) as the training objective function. Therefore, the loss function of speech enhancement \(\mathcal{L}_{Enh}\) can be defined as follows:
\begin{equation}
\mathcal{L}_{Enh}=\frac{1}{TF}\sum||\widetilde{M}\odot{|Y|}-|X|||_F^2
\label{eq5}
\end{equation}
where, \(\widetilde{M}\) is the estimated mask. The \(\odot\) denotes point-wise matrix multiplication. At the test stage, when the estimated mask \(|\widetilde{M}|\) is acquired by the trained enhancement network, we can estimate the enhanced magnitude spectrum \(|\widetilde{X}|\) by multiplying the \(|\widetilde{M}|\) the noisy spectrum \(|Y|\):
 \begin{equation}
 |\widetilde{X}|=\widetilde{M}\odot{|Y|}
 \label{eq512}
 \end{equation}

After the enhancement component, the Fbank features can be extracted from the STFT features by using the log Mel filterbank.

%
The enhanced and noisy Fbank features can be extracted as the follows:
	\begin{equation}
	\widetilde{O}_{enhanced}=log(Mel(|\widetilde{X}|))
	\label{eq6}
	\end{equation}
	\begin{equation}
	{O}_{noisy}=log(Mel(|Y|))
	\label{eq7}
	\end{equation}
	where \(\widetilde{O}_{enhanced}\) and \({O}_{noisy}\) denote the enhanced and noisy Fbank features, respectively. \(Mel(\cdot)\) is the operation of Mel matrix multiplication. From Eq.~\ref{eq6} and Eq.~\ref{eq7} we can know that the Fbank feature extraction procedure used as a layer of network is differentiable.

\subsection{Gated recurrent fusion network}

The conventional joint training method \cite{liu2019jointly} only uses the enhanced features for the speech recognition component, which is still affected by the speech distortion problem. In order to address this problem and extract more robust features for ASR, we apply the gated recurrent fusion algorithm \cite{liu20203d} to combine the noisy and enhanced features. Therefore, the raw fine structures from the noisy features can be learned to alleviate the speech distortion. In addition, noise signals can be also removed from the enhanced features.

As for the GRF network, we firstly use two parallel bidirectional long-short term memory (BLSTM) \(B(\cdot) \) networks to extract deep representations for enhanced (\(\widetilde{O}_{enhanced}\)) and noisy (\({O}_{noisy}\)) Fbank features, respectively. We denote them as \({\bm{\beta}}_{enhanced}\) (for enhanced representations) and \(\bm{\beta}_{noisy}\) (for noisy representations):
\begin{equation}
{\bm{\beta}}_{enhanced}=B(\widetilde{O}_{enhanced})
\label{eq61}
\end{equation}
\begin{equation}
\bm{\beta}_{noisy}=B({O}_{noisy})
\label{eq62}
\end{equation}

We apply the GRF to fuse the noisy and enhanced features. The gate structure in gated recurrent unit (GRU) \cite{cho2014learning} enables the selective fusion of multi-modal features. In this paper, we extend the GRU as our GRF block and modify it to fit the feature fusion.

Fig.~\ref{fig:grf} shows the overall schematic diagram of GRF and Fig.~\ref{fig:grf-block} shows the schematic diagram of GRF block. As shown in Fig.~\ref{fig:grf-block}, at first step (left), GRF block takes one of the noisy features \(\bm{\beta}_{noisy}\) as input. The outputs of this step will be used as the hidden state. Then, in the second step (right), GRF block applies the enhanced features \({\bm{\beta}}_{enhanced}\) as input. These two steps reuse the GRF block and share the same set of parameters. The GRF block consists of a reset gate, a update gate, a adaptive memory and a selective fusion. Next, we will utilize the first step with noisy feature \(\bm{\beta}_{noisy}\) as an example to explain in detail the principle and work flow of the GRF block.

At the first step of fusion between noisy features \(\bm{\beta}_{noisy}\) and enhanced features \({\bm{\beta}}_{enhanced}\), that is p = 1, we randomly initialize the hidden state \(h_0\). 

\textbf{Reset Gate} As for the reset gate, at step p, the hidden state \(h_p\) and the noisy input \(\bm{\beta}_{noisy}\) together to decide the status of the reset gate \(\textbf{\textsl{r}}\) by
\begin{equation}
\textbf{\textsl{r}}=\sigma(W_r,(\bm{\beta}_{noisy}, h_p))
\label{eq63}
\end{equation}
where \(\sigma\) denotes the sigmoid function, \(W_r\) is the weight of reset gate.

\textbf{Update Gate} The update gate \(\textbf{\textsl{z}}\) is also decided by the hidden state \(h_p\) and noisy input \(\bm{\beta}_{noisy}\). 
\begin{equation}
\textbf{\textsl{z}}=\sigma(W_z,(\bm{\beta}_{noisy}, h_p))
\label{eq64}
\end{equation}
where \(W_z\) is the weight of update gate.



\textbf{Adaptive Memory} As for adaptive memory, through the element-wise product \(\odot\), the reset gate \(\textbf{\textsl{r}}\) determines how much information in the past needs to be “memorized”.
\begin{equation}
h_p^{'}=\textbf{\textsl{r}}\odot{h_p}
\label{eq65}
\end{equation}
\begin{equation}
h_p^{c}=tanh(W_h(\bm{\beta}_{noisy}, h_p^{'}))
\label{eq66}
\end{equation}
where \(h_p^{c}\) acts similarly to the memory cell in the LSTM and helps the GRF block to remember long term information within the multi-stage fusion.

\textbf{Selective Fusion} The selective fusion aims to combine the noisy input \(\bm{\beta}_{noisy}\) and the hidden state \(h_p\). The fusion result at step \(p\) is
\begin{equation}
h_q=\textbf{\textsl{z}}\odot{h_p}+(1-\textbf{\textsl{z}})\odot{h_p^{c}}
\label{eq67}
\end{equation}
In this way, the forget gate \(\textbf{\textsl{z}}\) and the input gate \((1-\textbf{\textsl{z}})\) are linked. That is, if the previous information is ignored with a weight of \(\textbf{\textsl{z}}\), then the information for the current input \(h_p^{c}\) would be selected with a weight of \((1-\textbf{\textsl{z}})\).

For the one GRF stage, at step 1, we randomly initialize the hidden state. Then the noisy input \(\bm{\beta}_{noisy}\) is fed into the GRF block to acquire the output \(h_1\). Next, at step 2, the same structure and parameters are reused in the GRF block. The input hidden state is replaced by h1, and the input is replaced by the enhanced features \(\bm{\beta}_{enhanced}\), which is shown in Fig.~\ref{fig:grf-block}.

In this paper, we use the 4-stage GRF. After the 4-stage GRF, we can acquire the fusion features: \(f^{GRF}=h_p\). Finally, these fusion features \(f^{GRF}\), noisy and enhanced deep representations \(\bm{\beta}_{noisy}\) and \(\bm{\beta}_{enhanced}\) are used as the GRF features \(O_{GRF}\):
\begin{equation}
O_{GRF}=\rm Concat(\bm{\beta}_{noisy},\emph{f}^{GRF},\bm{\beta}_{enhanced})
\label{eq68}
\end{equation}
These GRF features \(O_{GRF}\) are used as the input of end-to-end ASR speech transformer.

\subsection{Speech transformer network}


In this paper, we apply the state-of-the-art end-to-end ASR method speech transformer with self-attention as our speech recognition component. In this subsection, the end-to-end speech recognition by speech transformer is described. 



Different from the RNNs, speech transformer utilizes the scaled dot-product attention to map the input sequence, which consists of three inputs queries \(Q\), keys \(K\), and values \(V\) with dimension \(d_{q}\), \(d_{k}\) and \(d_{v}\). The outputs of scaled dot-product attention can be defined as follows:
\begin{equation}
\rm Attention(Q,K,V)=\rm softmax(\frac{QK^T}{\sqrt{d_k}})V
\label{eq12}
\end{equation}


The multi-head attention (MHA) is defined as follows:
\begin{equation}
\rm MHA(Q,K,V)=\rm Concat(head_1,...,head_h)W_i^O
\label{eq13}
\end{equation}
\begin{equation}
\rm where\ head_i=Attention(QW_i^Q,KW_i^K,VW_i^V)
\label{eq14}
\end{equation}
Where the projections are parameter matrices \(W_i^Q \in{\mathbb{R}^{d_{m}\times{d_q}}},W_i^K \in{\mathbb{R}^{d_{m}\times{d_k}}},W_i^V \in{\mathbb{R}^{d_{m}\times{d_v}}}\) and \(W_i^O \in{\mathbb{R}^{hd_{v}\times{d_{m}}}}\). \(h\) is the number of heads.


The position-wise feed forward network (FFN) includes two linear transformations and a ReLU activation function in between.
\begin{equation}
FFN(x)=max(0,xW_1+b_1)W_2+b_2
\label{eq15}
\end{equation}
where \(x\), \(W_1\in{\mathbb{R}^{d_{m}\times{d_f}}}\), \(b_1\in{\mathbb{R}^{d_{m}}}\), \(W_2\in{\mathbb{R}^{d_{f}\times{d_m}}}\) and \(b_2\in{\mathbb{R}^{d_{m}}}\) are learnable parameters. In oder to make full use of the order of sequence, the sine and consine positional embedding are applied \cite{vaswani2017attention}.

Based on the cross-entropy criterion, the loss function of end-to-end speech recognition is defined as follows:
\begin{equation}
\mathcal{L}_{ASR}=-lnP(S^*|O_{GRF})
\label{eq16}
\end{equation}
where \(S^*\) is the ground truth of a whole sequence of output labels.

\subsection{Joint training}

The proposed system consists of three parts: the speech enhancement network, the gated recurrent fusion network and speech transformer network. By the joint training, these three parts are optimized simultaneously. It means that the parameters of these three parts are updated by stochastic gradient descent.


The loss function of the joint training is defined as follows:
\begin{equation}
\mathcal{L}=\mathcal{L}_{ASR}+\alpha\mathcal{L}_{Enh}
\label{eq17}
\end{equation}
Where the hyperparameter \(\alpha\) controls enhancement loss \(\mathcal{L}_{Enh}\).

\section{Experiments and Results}



\subsection{Dataset}

Our experiments are conducted on an open-source Mandarin speech corpus called AISHELL-1 \cite{bu2017aishell}. The AISHELL-1 consists of 400 speakers and over 170 hours of Mandarin speech data. The dataset includes three sets: training set, development set and test set. The training set contains about 150 hours with 120,098 utterances recorded by 340 speakers (161 males and 179 females). The development set contains about 10 hours with 14,326 utterance recorded by 40 speakers (12 males and 28 females) and test set contains about 5 hours with 7,176 utterances recorded by 20 speakers (13 males and 7 females).  For each speaker, around 360 utterances (about 26 minutes of speech) are released. Table~\ref{tab:dataset} shows the summary of all subsets in the AISHELL-1. The sampling rate of AISHELL-1 is 16000 Hz.

The noisy training set is generated by selecting each utterance from the AISHELL-1 training set. They are mixed with 100 Nonspeech Sounds noise database \footnote{http://web.cse.ohio-state.edu/pnl/corpus/HuNonspeech/HuCorpus.html} at signal-to-noise ratios (SNRs) randomly sampled between [0dB, 20dB]. For the test and development set, the 100 Nonspeech noises and NOISE-92 corpus noises \cite{varga1993assessment} are used to mix with the clean test set of AISHELL-1 with 5 different SNRs each 0dB, 5dB, 10dB, 15dB and 20dB.  We also test our model on the development set and the configurations of it are the same as the test set. Detailed configurations are listed in Table~\ref{tab:dataset_train}.

\begin{table}[t]
	\caption{The data structure of Mandarin speech corpus AISHELL-1. ``{hrs}`` means the hours of each set.}
	\label{tab:dataset}
	\centering
	\begin{tabular}{c|c|c|c|c}
		\toprule
		Subset      & Duration(hrs) & \#Sentences & \#Male & \#Female \\ \midrule
		Training    & 150           & 120098      & 161    & 179      \\ 
		Development & 10            & 14326       & 12     & 28       \\ 
		Test        & 5             & 7200        & 13     & 7        \\ \bottomrule
	\end{tabular}
\end{table}

\begin{table}[t]
	\caption{The data structure of our training set for the speech enhancement, MCT and the joint training methods.}
	\label{tab:dataset_train}
	\centering
	\begin{tabular}{c|c|c}
		\toprule
		Subset & SNRs                                                                                & Noise corpus                                                                 \\ \midrule
		Training & \begin{tabular}[c]{@{}c@{}}randomly sampled \\ between {[}0dB, 20dB{]}\end{tabular} & 100 Nonspeech Sounds                                                         \\ \midrule
		Development     & 0, 5, 10, 15 and 20dB                                                               & \begin{tabular}[c]{@{}c@{}}100 Nonspeech Sounds \\ and NOISE-92\end{tabular} \\ \midrule
		Test     & 0, 5, 10, 15 and 20dB                                                               & \begin{tabular}[c]{@{}c@{}}100 Nonspeech Sounds \\ and NOISE-92\end{tabular} \\ \bottomrule
	\end{tabular}
\end{table}

\subsection{Experimental setup}

For speech enhancement network, the 257-dim amplitude spectrum of the noisy speech are used as input features, which are computed by using a STFT with 32 ms length hamming window and 16 ms window shift. To acquire a quite better enhancement performance, three BLSTM layers with 512 nodes are used as the speech enhancement network. And a linear layer with the ReLU activation function is connected to the last BLSTM layer (mask layer), whose output size is equal to the input size. The aim of the ReLU activation function is to estimate the amplitude mask of the clean target speech. In addition, the enhancement network contains random dropouts with a dropout rate 0.5.

The Fbank extraction network is a linear layer to convert the STFT features to Fbank features, which operates \(257\times80\) matrix multiplication. After the matrix multiplication, we also do the logarithmic operation. The dimension of the Fbank features is 80. From the Eq.~\ref{eq6} and Eq.~\ref{eq7} we can know that the Fbank feature extraction procedure used as a layer of network is differentiable. Therefore, the enhancement network and end-to-end ASR can be jointly trained.

For gated recurrent fusion network, we firstly utilize two parallel BLSTM networks with two layers and 320 nodes to extract \({\bm{\beta}}_{enhanced}\) (for enhanced representations) and \({\bm{\beta}}_{noisy}\) (for noisy representations). The GRF network also contains random dropouts with a dropout rate 0.5. Finally, a linear layer with the ReLU activation function is connected to the GRF representations \({O}_{GRF}\), whose output size is 320. Therefore, the dimension of the GRF representations \({O}_{GRF}\) is 320.

For speech transformer network, we use 6 self-attention blocks as encoder and 6 self-attention blocks as prediction network. We take \((d_k,h,d_f)=(512,4,1024)\). Before the Transformer encoder, the GRF representations \({O}_{GRF}\) are encoded by two convolutional neural network (CNN) blocks. The CNN layers have a kernel size of \(3\times3\). We chose 4232 characters (including ’blank’ and ’unk’ labels) as output labels.

As for the SpecAugment, we set the time mask with 15 and frequency mask with 27.

All our models optimized in the same method as \cite{salazar2019self}. And we set warm up steps to 8000 and the factor of learning rate to 0.5. During decoding, we use beam search with width of 5 for all the experiments. Our models are implemented using Pytorch deep learning framework \footnote{Available online at https://pytorch.org/}.

Before joint training the whole model, the enhancement component is initialized by the speech enhancement network and the ASR model is initialized by the model trained by clean data.


\begin{table}[t]
	\caption{The annotations of different models.}
	\label{tab:model_name}
	\centering
	\begin{tabular}{p{2cm}p{5.5cm}}
		\toprule
		Model name & Annotations   \\ \midrule
		E2E\_ASR-clean & Speech transformer system trained by clean data.                                                         \\ \midrule
		E2E\_ASR-SpecAugment     & Speech transformer system trained by SpecAugment algorithm. \\ \midrule
		E2E\_ASR-MCT      & Speech transformer system trained by multi-condition training (MCT) algorithm.  \\ \midrule
		E2E\_ASR-MCT-SpecAugment     & Speech transformer system trained by SpecAugment and MCT algorithm, simultaneously \\ \midrule
		Joint-Enhance-E2E\_ASR     & The model is trained by applying the conventional joint training method, which only uses the enhanced features as the input of speech transformer. \\ \midrule
		Joint-Enhance-E2E\_ASR-concat     & The model is trained by the joint training framework, which uses the concatenated \({\bm{\beta}}_{enhanced}\) and \({\bm{\beta}}_{noisy}\) as the input of speech transformer. \\ \midrule
		Joint-Enhance-E2E\_ASR-GRF     & The model is trained by the joint training framework, which applies the GRF to fuse the noisy and enhanced features. \\ \bottomrule
	\end{tabular}
\end{table}

\begin{table*}[!ht]
	\caption{CER results of speech transformer system trained by clean data, SpecAugment and multi-condition training (MCT) without the enhancement on the test set.}
	\label{tab:results1}
	\renewcommand\tabcolsep{13pt}
	\centering
	\begin{tabular}{c|cccccc|c}
		\toprule
		\multirow{2}{*}{Speech transformer} & \multicolumn{7}{c}{CER Results(\%) on Test Set}                   \\ \cline{2-8} 
		 & 0dB    & 5dB   & 10dB  & 15dB  & 20dB  & AVG. & clean \\ \midrule
		E2E\_ASR-clean                    & 107.03 & 72.71 & 45.49 & 26.37 & 18.01 & 53.33 & 10.74 \\ \midrule
		E2E\_ASR-SpecAugment       & 68.44 & 43.74 & 25.89 & 14.97 & 11.59 & 32.55 & 8.78 \\ \midrule
		E2E\_ASR-MCT                       & 27.92  & 19.13 & 14.72 & 11.96 & 10.36 & 16.71 & 9.64 \\  \midrule
		E2E\_ASR-MCT-SpecAugment                       & 24.43  & 16.85 & 14.11 & 11.16 & 9.66 & 15.14 & 9.30 \\  \bottomrule
	\end{tabular}
\end{table*}

\begin{table*}[!ht]
	\caption{CER results of speech transformer system trained by clean data, SpecAugment and MCT without the enhancement on the development set.}
	\label{tab:results11}
	\renewcommand\tabcolsep{13pt}
	\centering
	\begin{tabular}{c|cccccc|c}
		\toprule
		\multirow{2}{*}{Speech transformer} & \multicolumn{7}{c}{CER Results(\%) on Development Set}                   \\ \cline{2-8} 
		 & 0dB    & 5dB   & 10dB  & 15dB  & 20dB  & AVG. & clean  \\ \midrule
		E2E\_ASR-clean                    & 98.30 & 62.64& 37.15 & 21.81 & 14.76 & 46.81 & 9.10 \\ \midrule
		E2E\_ASR-SpecAugment       & 63.01 & 38.07 & 21.56 & 13.51 & 10.36 & 29.24 & 7.96 \\ \midrule
		E2E\_ASR-MCT                       & 25.26  & 16.11 & 12.16 & 10.09 & 9.22 &14.55 & 8.51 \\  \midrule
		E2E\_ASR-MCT-SpecAugment           & 21.76  & 14.44 & 11.13 & 9.92 & 9.24 & 13.28 & 8.44 \\  \bottomrule
	\end{tabular}
\end{table*}

\begin{table*}[!ht]
	\caption{CER results of speech transformer system trained by clean data, SpecAugment and MCT on the enhanced test set. The clean and noisy test data all are enhanced by the speech enhancement model to remove the noise signals.}
	\label{tab:results2}
	\renewcommand\tabcolsep{13pt}
	\centering
	\begin{tabular}{c|cccccc|c}
		\toprule
		\multirow{2}{*}{Speech transformer} & \multicolumn{7}{c}{CER Results(\%) on Enhanced Test Set}                  \\ \cline{2-8} 
		& 0dB   & 5dB   & 10dB  & 15dB  & 20dB  & AVG. & clean  \\ \midrule
		E2E\_ASR-clean                    & 44.11 & 29.85 & 20.12 & 14.75 & 12.51 & 24.06 & 14.23 \\ \midrule
		E2E\_ASR-SpecAugment       & 34.82 & 22.22 & 15.50 & 11.45 & 10.02 & 18.64 & 11.21 \\ \midrule
		E2E\_ASR-MCT                      & 28.74 & 18.82 & 14.18 & 11.63 & 10.37 & 16.63 & 12.73 \\  \midrule
		E2E\_ASR-MCT-SpecAugment           & 27.04  & 18.01 & 13.88 & 10.80 & 9.56 & 15.74 & 11.38 \\  \bottomrule
	\end{tabular}
\end{table*}

\begin{table*}[!ht]
	\caption{CER results of speech transformer system trained by clean data, SpecAugment and MCT on the enhanced development set.}
	\label{tab:results21}
	\renewcommand\tabcolsep{13pt}
	\centering
	\begin{tabular}{c|cccccc|c}
		\toprule
		\multirow{2}{*}{Speech transformer} & \multicolumn{7}{c}{CER Results(\%) on Enhanced Development Set}                  \\ \cline{2-8} 
		 & 0dB   & 5dB   & 10dB  & 15dB  & 20dB  & AVG. & clean \\ \midrule
		E2E\_ASR-clean                   & 37.81 & 24.13 & 16.36 & 12.34 & 10.60 & 20.21  & 11.84 \\ \midrule
		E2E\_ASR-SpecAugment      & 30.62& 18.87 & 12.82 & 10.19 & 9.15 & 16.31 & 9.95 \\ \midrule
		E2E\_ASR-MCT                      & 24.32 & 15.52 & 11.34 & 9.95 & 9.00 & 14.01 & 10.99 \\  \midrule
		E2E\_ASR-MCT-SpecAugment           & 23.37  & 14.92 & 11.21 & 9.77 & 9.17 & 13.67 & 10.13 \\  \bottomrule
	\end{tabular}
\end{table*}

In the following results, the character error rate (CER) is applied to quantify the system performance. "clean" refers to the original clean AISHELL-1 test and development set. "0dB, 5dB, 10dB, 15dB and 20dB" denote the different SNR conditions of the test and development set. "AVG." means the average result of these SNRs.

Speech transformer method is more powerful to model long-term dependencies than RNNs-based sequence to sequence models. In addition, it is the state-of-the-art method for end-to-end speech recognition. Therefore, we apply the speech transformer as end-to-end ASR model for all of the speech recognition component.

\subsection{Evaluation of the end-to-end ASR without the enhancement}

Table~\ref{tab:results1} and Table~\ref{tab:results11} show the results of speech transformer trained by clean data, SpecAugment and MCT on the test and development set, respectively. We use the clean and noisy test data (without enhancement for the test and development data) to test the ASR models. The E2E\_ASR-clean means that only the clean data is applied to train the speech transformer model. The E2E\_ASR-SpecAugment denotes that the speech transformer model is trained by SpecAugment algorithm. And the E2E\_ASR-MCT denotes that the speech transformer model is trained by the multi-condition training method, which uses both the clean and generated noisy speech data. The E2E\_ASR-MCT-SpecAugment denotes that the speech transformer model is trained by the MCT and SpecAugment algorithm, simultaneously. Table~\ref{tab:model_name} shows the annotations of different models.

\begin{table*}[!ht]
	\caption{CER results of speech transformer system trained by clean data, SpecAugment and MCT on the test set with SNR to fuse the original and enhanced data.}
	\label{tab:results_snr1}
	\renewcommand\tabcolsep{13pt}
	\centering
	\begin{tabular}{c|cccccc|c}
		\toprule
		\multirow{2}{*}{Speech transformer} & \multicolumn{7}{c}{CER Results(\%) on Test Set with SNR Data}                  \\ \cline{2-8} 
		& 0dB   & 5dB   & 10dB  & 15dB  & 20dB  & AVG. & clean  \\ \midrule
		E2E\_ASR-clean                    & 99.91 & 68.33 & 42.82 & 24.84 & 17.39 & 50.11 & 10.73 \\ \midrule
		E2E\_ASR-SpecAugment       & 64.41 & 40.44 & 23.97 & 14.36 & 11.30 & 30.55 & 8.73 \\ \midrule
		E2E\_ASR-MCT              & 26.44& 18.44 & 14.41 & 11.77 & 10.32 & 16.17   & 9.68     \\  \midrule
		E2E\_ASR-MCT-SpecAugment           & 23.25  & 16.34 & 13.77 & 11.00 & 9.57 & 14.70 & 9.34 \\  \bottomrule
	\end{tabular}
\end{table*}

\begin{table*}[!ht]
	\caption{CER results of speech transformer system trained by clean data, SpecAugment and MCT on the development set with SNR to fuse the original and enhanced data.}
	\label{tab:results_snr2}
	\renewcommand\tabcolsep{13pt}
	\centering
	\begin{tabular}{c|cccccc|c}
		\toprule
		\multirow{2}{*}{Speech transformer} & \multicolumn{7}{c}{CER Results(\%) on Development Set with SNR Data}                  \\ \cline{2-8} 
		& 0dB   & 5dB   & 10dB  & 15dB  & 20dB  & AVG. & clean \\ \midrule
		E2E\_ASR-clean                    & 92.28 & 57.44 & 34.47 & 20.61 & 14.06 &43.67  & 9.12\\ \midrule
		E2E\_ASR-SpecAugment       & 58.74 & 35.17 & 20.01 & 12.89 & 10.25 & 27.35 & 7.96 \\ \midrule
		E2E\_ASR-MCT              & 24.01& 15.50 & 11.94 & 10.04 & 9.17 & 14.11    & 8.50    \\  \midrule
		E2E\_ASR-MCT-SpecAugment           & 20.62  & 13.93 & 10.99 & 9.89 & 9.20 & 12.91 & 8.44 \\  \bottomrule
	\end{tabular}
\end{table*}

From Table~\ref{tab:results1} and Table~\ref{tab:results11} we can find that on the noisy condition, the E2E\_ASR-clean performs poorly, which demonstrates the necessity of the robust end-to-end ASR investigation. Although E2E\_ASR-SpecAugment can acquire a better performance on the clean condition, it is still performs poorly on the noisy condition. However, as for the E2E\_ASR-MCT, it significantly improves the system robustness. Compared with the E2E\_ASR-clean, the E2E\_ASR-MCT obtains 68.67\% relative reduction on the noisy "AVG." condition. Especially on the low SNR condition, which means the noisy speech is dominated by the noise signals, the E2E\_ASR-MCT can acquire significantly better results than the E2E\_ASR-clean. This is because that the multi-condition training method is trained by both the clean and noisy speech data, which masks the E2E\_ASR-MCT more robust than the E2E\_ASR-clean. In addition, the SpecAugment algorithm can improve the performance of ASR on the clean condition. However, on the noisy condition, the performance of it is worse than the E2E\_ASR-MCT. When the MCT and SpecAugment algorithm are applied simultaneously, we can find that the performance of ASR can be improved.


In addition, from Table~\ref{tab:results1} and Table~\ref{tab:results11}  we can also find that the E2E\_ASR-MCT is even better than E2E\_ASR-clean on the ``clean`` test set condition. Compared with the E2E\_ASR-clean, the E2E\_ASR-MCT obtains 5.21\% relative reduction. The reason is that although the test set of AISHELL-1 is quite clean, the recorded speech signals may consist of slight noise because of the background environment. The E2E\_ASR-MCT model contains both various noise distributions and clean data distributions. Therefore, the performance of E2E\_ASR-MCT is better than the E2E\_ASR-clean for the clean test set.

\subsection{Evaluation of the effect of speech enhancement for end-to-end ASR}

According to the Section~\ref{sec:enhancement}, we train the mask-based enhancement network, which aims to remove the noise signals. The enhanced features in test and development sets are fed into the well-trained E2E\_ASR-clean and E2E\_ASR-MCT network to predict the final label sequence. These results are shown in the Table~\ref{tab:results2} and Table~\ref{tab:results21}. 

From Table~\ref{tab:results2} and Table~\ref{tab:results21} we can make several observations. Firstly, the speech enhancement component can significantly improve the system robustness for E2E\_ASR-clean, which confirms the effectiveness of combining the speech enhancement with end-to-end ASR framework. Compared with the without enhancement condition, the speech enhancement can obtain 54.88\% relative reduction for E2E\_ASR-clean on the noisy "AVG." test set condition. The reason is that the enhanced speech can removed most of the noise signals so that the enhancement can reduce the influences of noise. As for this condition, although the speech distortion still damages the performance of ASR, the noise signals are more harmful to the speech recognition. Therefore, the speech enhancement can improve the performance of E2E\_ASR-clean.

Secondly, as for the E2E\_ASR-MCT, the performance of enhancement is comparable to the method of without enhancement condition for the noisy test set. For example, as for the test set, the enhancement CER of E2E\_ASR-MCT is 16.63\% on the noisy "AVG." condition and the without enhancement CER is 16.71\%. This is because that the E2E\_ASR-MCT is more robust than E2E\_ASR-clean. The denoising seems not as important as E2E\_ASR-clean. On this condition, the improvement of denoising makes up the damage of speech distortion. In addition, the E2E\_ASR-MCT-SpecAugment on the enhanced test set have a higher WER than on the non-enhanced test set. This is because that the E2E\_ASR-MCT-SpecAugment is more robust than the other methods for the noisy data. When we use the enhanced data to test models, the denoising is not so important for the E2E ASR-MCT-SpecAugment, but the speech distortion has damage for it.

\begin{table*}[!ht]
	\caption{CER results of speech transformer system using the joint training method on the test set.}
	\label{tab:results3}
	\renewcommand\tabcolsep{11pt}
	\centering
	\begin{tabular}{c|cccccc|c}
		\toprule
		\multirow{2}{*}{Models}                              & \multicolumn{7}{c}{CER Results(\%) on Test Set}                                                                                                                                                                                                                                   \\ \cline{2-8} 
		                              & 0dB                                 & 5dB                                 & 10dB                                & 15dB                                & 20dB                                & AVG.          & clean                      \\ \midrule
		Joint-Enhance-E2E\_ASR                                                         & 25.17                               & 17.87                              & 14.18                               & 11.83                               & 10.63                               & 15.84                & 10.18               \\ \midrule
		Joint-Enhance-E2E\_ASR-concat                                            & 23.65                               & 16.63                               & 13.37                               & 11.19                              & 10.15                              & 14.91          & 9.72                      \\ \midrule
		Joint-Enhance-E2E\_ASR-GRF (our)         & \textbf{21.98} & \textbf{15.87} & \textbf{13.09} & \textbf{10.70} & \textbf{9.98} & \textbf{14.25}  & \textbf{9.48} \\ \bottomrule
	\end{tabular}
\end{table*}

\begin{table*}[!ht]
	\caption{CER results of speech transformer system using the joint training method on the development set.}
	\label{tab:results31}
	\renewcommand\tabcolsep{11pt}
	\centering
	\begin{tabular}{c|cccccc|c}
		\toprule
		\multirow{2}{*}{Models}                              & \multicolumn{7}{c}{CER Results(\%) on Development Set}                                                                                                                                                                                                                                   \\ \cline{2-8} 
		                              & 0dB                                 & 5dB                                 & 10dB                                & 15dB                                & 20dB                                & AVG.     & clean                           \\ \midrule
		Joint-Enhance-E2E\_ASR                                                        & 22.14                               & 14.42                              & 11.70                               & 10.27                               & 9.54                              & 13.60               & 8.88                \\ \midrule
		Joint-Enhance-E2E\_ASR-concat                                             & 20.61                              & 13.35                              & 10.59                               & 9.40                              & 8.94                              & 12.56                & 8.45               \\ \midrule
		Joint-Enhance-E2E\_ASR-GRF (our)          & \textbf{19.76} & \textbf{12.71} & \textbf{9.93} & \textbf{9.09} & \textbf{8.74} & \textbf{12.03} & \textbf{8.18} \\ \bottomrule
	\end{tabular}
\end{table*}

Finally, for the clean test set condition, the performances of both E2E\_ASR-clean,  E2E\_ASR-SpecAugment  and E2E\_ASR-MCT are degraded after the speech enhancement. Specifically, after the speech enhancement, the E2E\_ASR-clean CER increases from 10.74\% to 14.23\% on the noisy "AVG." condition. For the E2E\_ASR-MCT, it increases from 9.64\% to 12.73\%. Because the SNR of clean test set is very high, which means that the speech includes very little even no noise. So the denoising is unnecessary. However, the speech distortion could dramatically degrade the speech recognition performance. Therefore, after the speech enhancement, end-to-end ASR performances are damaged for the clean test set. These results confirm the importance of addressing the speech distortion problem. In addition, the enhancement and end-to-end ASR network are separately trained by the different objectives and the enhanced process may introduce unseen distortions that degrades the performance. This is the reason that we propose the gated recurrent fusion features with the joint enhancement and speech transformer  training method to improve the robustness of end-to-end ASR.


\subsection{Evaluation of the SNR test data between the original and enhanced data}

Table~\ref{tab:results_snr1} and Table~\ref{tab:results_snr2} show the oracle experimental results that apply the SNR to fuse the original and enhanced data. 
\begin{equation}
f_{SNR}=20log_{10}(\frac{y}{\widetilde{y}})
\label{eqsnr}
\end{equation}
where \(y\) is the original data and \(\widetilde{y}\) is the enhanced data, \(f_{SNR}\) denotes the SNR fusion data. We set the SNR 20 dB (between the original and enhanced data). From Table~\ref{tab:results_snr1} and Table~\ref{tab:results_snr2} we can find that when the SNR data is applied, all the ASR performances are improved on the clean condition. These results indicate that the SNR between the original and enhanced data can alleviate the speech distortion in some degree. However, on the noisy condition, most of the performances of ASR models are worse than the only enhanced test data (compared with Table~\ref{tab:results2} and Table~\ref{tab:results21}). This is because that the SNR test data can produce noise to the waves, which is harmful to ASR. 

As for the E2E\_ASR-MCT-SpecAugment, compared with Table~\ref{tab:results2} and Table~\ref{tab:results21}, we can find that it's performances can be improved on both clean and noisy conditions. Specifically, when the SNR test data is applied, the E2E\_ASR-MCT-SpecAugment increases from 11.38\% to 9.34\% and 15.74\% to 14.70\% on clean and noisy "AVG." conditions. These results show the effectiveness of E2E\_ASR-MCT-SpecAugment method.

\subsection{Evaluation of the joint enhancement and speech transformer training method for robustness end-to-end ASR}

In order to address the speech distortion problem, we propose to apply the joint enhancement and speech transformer training method with gated recurrent fusion features for robust end-to-end ASR. Table~\ref{tab:results3} and Table~\ref{tab:results31} show the CER results of speech transformer system using the joint training method on the test and development sets, respectively. The ``{Joint-Enhance-E2E\_ASR}" means applying the conventional joint training method, which only uses the enhanced features as the input of speech transformer. The ``{Joint-Enhance-E2E\_ASR-concat}" means using the concatenated \({\bm{\beta}}_{enhanced}\) and \({\bm{\beta}}_{noisy}\) as the input of speech transformer. ``{Joint-Enhance-E2E\_ASR-GRF}" is our proposed method, which utilizes the GRF representations \({O}_{GRF}\) as nput of speech transformer. For all of the joint training methods, we set the hyperparameter of enhancement loss \(\mathcal{L}_{Enh}\) \(\alpha=1.0\) in Eq.~\ref{eq17}.

From Table~\ref{tab:results3} and Table~\ref{tab:results31} we can make several observations. Firstly, the joint enhancement and speech transformer training method can improve the performance of end-to-end ASR, which demonstrates the effectiveness of the joint training method. For example, compared with the E2E\_ASR-MCT with enhancement method (in Table~\ref{tab:results2}), the Joint-Enhance-E2E\_ASR method can reduce the CER from 16.63\% to 15.84\%  on the noisy "AVG." condition. Especially on the low SNR=0 condition, the CER can be further reduced from 28.74\% to 25.17\%. The Joint-Enhance-E2E\_ASR method simultaneously optimizes the speech enhancement and speech recognition by the joint training algorithm. This is because that speech enhancement and speech recognition are not two independent tasks and they can clearly benefit from each other. Therefore, the Joint-Enhance-E2E\_ASR method can acquire better results on the noisy conditions than E2E\_ASR-MCT.

Secondly, although the Joint-Enhance-E2E\_ASR method can improve the robustness of end-to-end ASR, it degrades the system performance for the clean test set compared with the E2E\_ASR-MCT without enhancement (in Table~\ref{tab:results1}). It increases the CER from 9.64\% to 10.18\% with 5.30\% relative increase. This indicates that only using the enhanced features as the input of speech transformer for joint training method can also be affected by the speech distortion problem in some degree. Therefore, the performance of the Joint-Enhance-E2E\_ASR method for the clean test set is worse than the E2E\_ASR-MCT without enhancement.


Thirdly, when the noisy and enhanced features are fused by a deep method, the end-to-end ASR performance of joint training method can be further improved. Instead of using the gated recurrent fusion features, we fuse the deep noisy \({\bm{\beta}}_{noisy}\) and enhanced features \({\bm{\beta}}_{enhanced}\) by concatenate operation: 
\begin{equation}
\widetilde{O}_{concat}=\{{\bm{\beta}}_{enhanced};{\bm{\beta}}_{noisy}\}
\label{eq18}
\end{equation}
Then the \(\widetilde{O}_{concat}\) is used as the input of speech transformer. This method is denoted as Joint-Enhance-E2E\_ASR-concat. Compared with the conventional Joint-Enhance-E2E\_ASR method, the Joint-Enhance-E2E\_ASR-concat obtains 5.87\% and 7.65\% relative reduction on "AVG." test and development set condition. In addition, for the clean test set, the performance of Joint-Enhance-E2E\_ASR-concat is comparable to the E2E\_ASR-MCT method. These results indicate that combining the noisy and enhanced features can not only boost the robustness of end-to-end ASR, but also alleviate the speech distortion problem.

Finally, our proposed gated recurrent fusion algorithm with joint enhancement and speech transformer training method Joint-Enhance-E2E\_ASR-GRF can significantly improve the performance of end-to-end speech recognition. Specifically, the performance of Joint-Enhance-E2E\_ASR-GRF all is better than Joint-Enhance-E2E\_ASR-concat  method no matter what conditions. In addition, compared with the E2E\_ASR-MCT without enhancement method, our proposed Joint-Enhance-E2E\_ASR-GRF method obtains 14.72\% relative reduction on "AVG." test set condition. Note that, our proposed method can obtain a better performance in the low SNRs. For example, when SNR is 0 dB, Joint-Enhance-E2E\_ASR-GRF can reduce the CER from 27.92\% to 21.98\% with 21.28\% relative reduction. Compared with the Joint-Enhance-E2E\_ASR method, our proposed Joint-Enhance-E2E\_ASR-GRF method acquires 10.04\% and 11.54\% relative reduction on "AVG." test and development set condition. These results suggest the potential of our proposed method. From the clean test condition results we can find that when the gated recurrent fusion features are used, the performance of end-to-end speech recognition even gets the better performance than the E2E\_ASR-clean and E2E\_ASR-MCT methods. This result indicates that our proposed method could alleviate the speech distortion problem very well.

\subsection{Evaluation of the model parameters}

Table~\ref{tab:results_parameter} shows the number of parameters of different models. "Speech Enhancement" means the component parameters of speech enhancement. "Speech Transformer" denotes the parameters of standard speech transformer model. "Speech Transformer(concat)" is the ASR component parameters of Joint-Enhance-E2E\_ASR-concat model. "Speech Transformer(GRF)" means the ASR component parameters of our proposed Joint-Enhance-E2E\_ASR-GRF model.

From Table~\ref{tab:results_parameter}, we can find that the number of parameters of our proposed model are more than the standard speech transformer model. The main reason is that our proposed model has another BLSTM with two layers, which lead to more parameters. In addition, compared with the "Speech Transformer(concat)", the "Speech Transformer(GRF)" has more about 1.02 millions parameters. These parameters mainly come from the GRF algorithm.

In order to discuss the computational complexity of the GRF algorithm, we also compute the training time of Joint-Enhance-E2E\_ASR-concat and Joint-Enhance-E2E\_ASR-GRF. Specifically, the training time of each utterance of Joint-Enhance-E2E\_ASR-concat is 0.622s and the Joint-Enhance-E2E\_ASR-GRF is 0.746s. Therefore, compared with the Joint-Enhance-E2E\_ASR-concat, the Joint-Enhance-E2E\_ASR-GRF has more 1.02 millions parameters and 0.124s training time, which indicates that the computational complexity of GRF algorithm is not so high.

\begin{table}[t]
	\centering
	\caption{The number of parameters of different models.}
	\label{tab:results_parameter}
	\begin{tabular}{|c|c|}
		\hline
		Model                       & Parameters(Million) \\ \hline
		Speech Enhancement          & 16.02          \\ \hline
		Speech Transformer          & 25.40          \\ \hline
		Speech Transformer(concat) & 39.76          \\ \hline
		Speech Transformer(GRF)     & 40.78          \\ \hline
	\end{tabular}
\end{table}

\section{Discussions}

The above experimental results show that our proposed joint enhancement and speech transformer method with gated recurrent fusion features could boost the robustness of end-to-end speech recognition and obtain a quite good performance. In addition, it could alleviate the speech distortion problem very well. Some interesting observations can be made as follows. 

On the noisy condition, the joint enhancement and speech transformer method can obtain a better performance for end-to-end robust speech recognition than the E2E\_ASR-SpecAugment and E2E\_ASR-MCT methods. E2E\_ASR-SpecAugment and E2E\_ASR-MCT only apply the end-to-end ASR loss function to optimize the model, which deal with the speech enhancement and speech recognition as two independent tasks. These methods are affected by the speech distortion problem and highly dependent on the performance of the enhancement front-end. However, the speech enhancement and speech recognition are not two independent tasks and they can clearly benefit from each other. Our proposed Joint-Enhance-E2E\_ASR method simultaneously optimizes the enhancement and speech recognition by the joint training algorithm. Therefore, the Joint-Enhance-E2E\_ASR is more robust than E2E\_ASR-MCT. In addition, the Joint-Enhance-E2E\_ASR could alleviate the speech distortion problem in some degree.

These gated recurrent fusion representations of enhanced and noisy features are effective to address the speech distortion problem. From Fig.~\ref{fig:yuputu} we can find that because of the leaks by the enhancement network, so much of important speech information is lost. The GRF aims to offset these leaks by fusing the noisy and enhanced data. Compared with the Joint-Enhance-E2E\_ASR method, which only uses the enhanced features as the input of end-to-end ASR, our proposed Joint-Enhance-E2E\_ASR-GRF method can acquire a better performance. In addition, our proposed Joint-Enhance-E2E\_ASR-GRF method even obtains better result on the clean condition than the E2E\_ASR-clean and E2E\_ASR-MCT methods. The reason is that these GRF representations are extracted by the gated recurrent module, which can dynamically fuse the enhanced and noisy features. Therefore, these GRF representations can learn the raw fine structures from the noisy features so that they can alleviate the speech distortion. Meanwhile, they can also remove the noise signals from the enhanced features. So they are more appropriate and robust features for the end-to-end speech recognition.

In summary, our proposed the joint enhancement and speech transformer training method with gated recurrent fusion representations can effectively boost the robustness of end-to-end speech recognition. In addition, it can address the speech distortion problem very well.

\section{Conclusion}

In this paper, we propose a joint enhancement and speech transformer training method with gated recurrent fusion for robust end-to-end speech recognition. The joint training compositional scheme is used to simultaneously optimize the enhancement and speech recognition. In addition, in order to address the speech distortion problem and extract more robust features for end-to-end ASR, we apply the gated recurrent fusion algorithm to combine the noisy and enhanced features. Experiments on Mandarin AISHELL-1 demonstrate that our proposed method is effective for the robust end-to-end ASR and can solve the speech distortion problem very well. In future, we will explore the time domain speech enhancement to acquire a better enhanced speech and obtain greater performance improvement for our proposed method.

%

\ifCLASSOPTIONcaptionsoff
  \newpage
\fi



%

%
%

\bibliography{references}{}
\bibliographystyle{IEEEtran}

%

\begin{IEEEbiography}[{\includegraphics[width=1.1in,height=1.25in,clip,keepaspectratio]{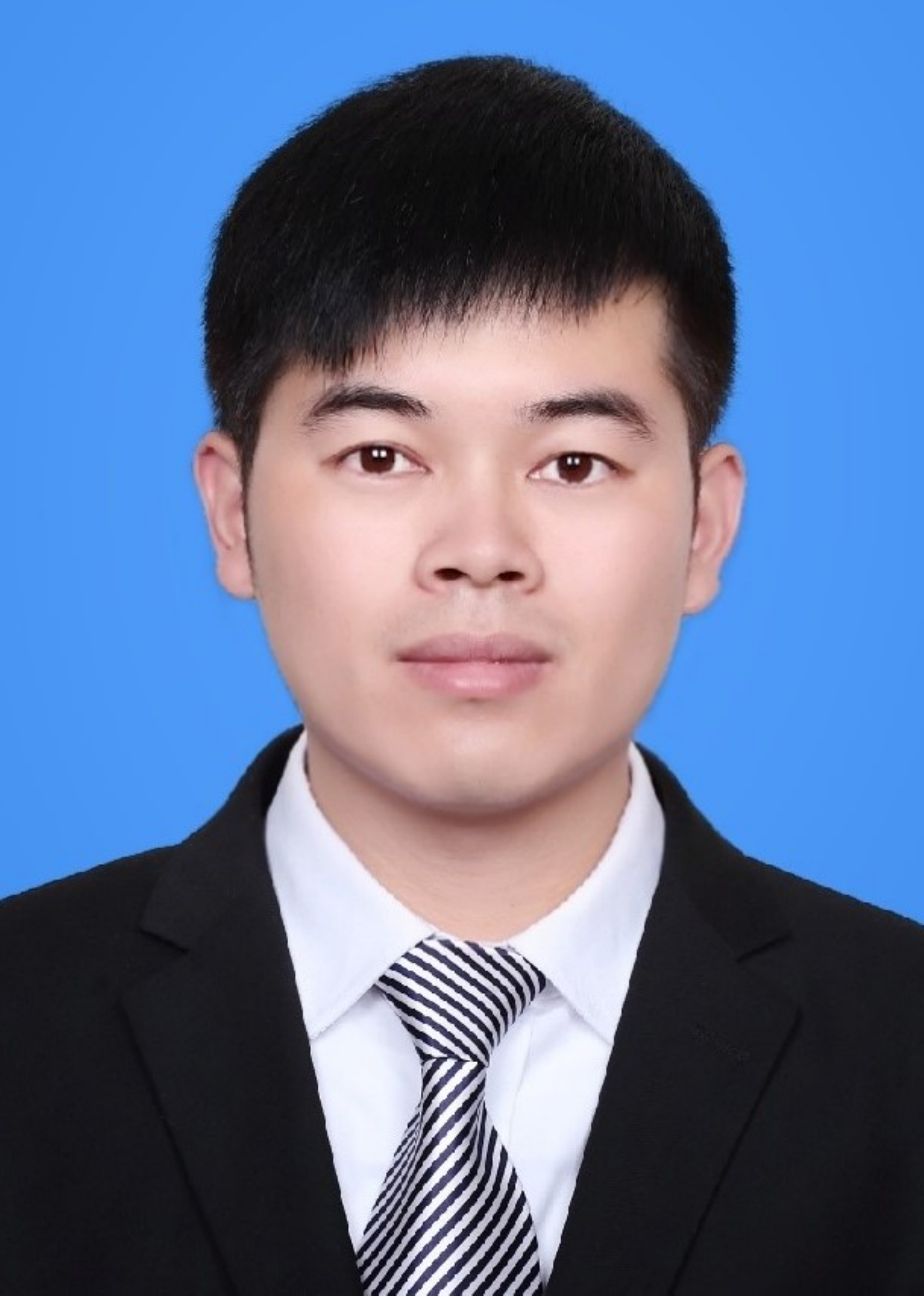}}]{Cunhang Fan}
	received the B.S. degree from the Beijing University of Chemical Technology (BUCT), Beijing, China, in 2016. He is currently working toward the Ph.D degree with the National Laboratory of Pattern Recognition (NLPR), Institute of Automation, Chinese Academy of Sciences (CASIA), Beijing, China. His current research interests include speech enhancement, speech recognition and speech signal processing.
\end{IEEEbiography}
\begin{IEEEbiography}[{\includegraphics[width=1.1in,height=1.25in,clip,keepaspectratio]{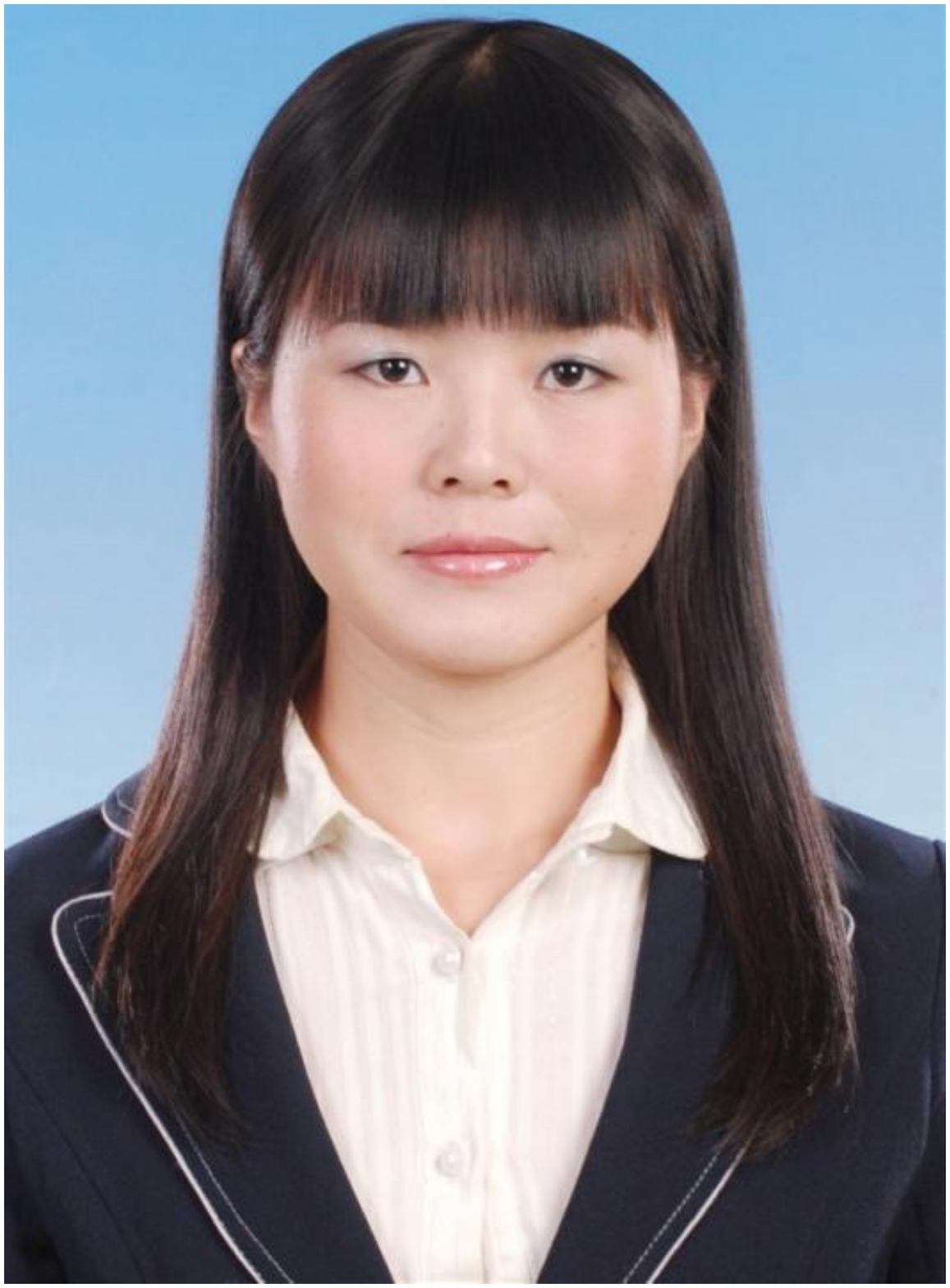}}]{Jiangyan Yi}
	received the Ph.D. degree from the University of Chinese Academy of Sciences, Beijing, China, in 2018, and the M.A. degree fromthe Graduate School of Chinese Academy of Social Sciences, Beijing, China, in 2010. She was a Senior R\&D Engineer with Alibaba Group during 2011 to 2014. She is currently an Assistant Professor with the National Laboratory of Pattern Recognition, Institute of Automation, Chinese Academy of Sciences, Beijing, China. Her current research interests include speech processing, speech recognition, distributed computing, deep learning, and transfer learning. 
\end{IEEEbiography}
\begin{IEEEbiography}[{\includegraphics[width=1.1in,height=1.25in,clip,keepaspectratio]{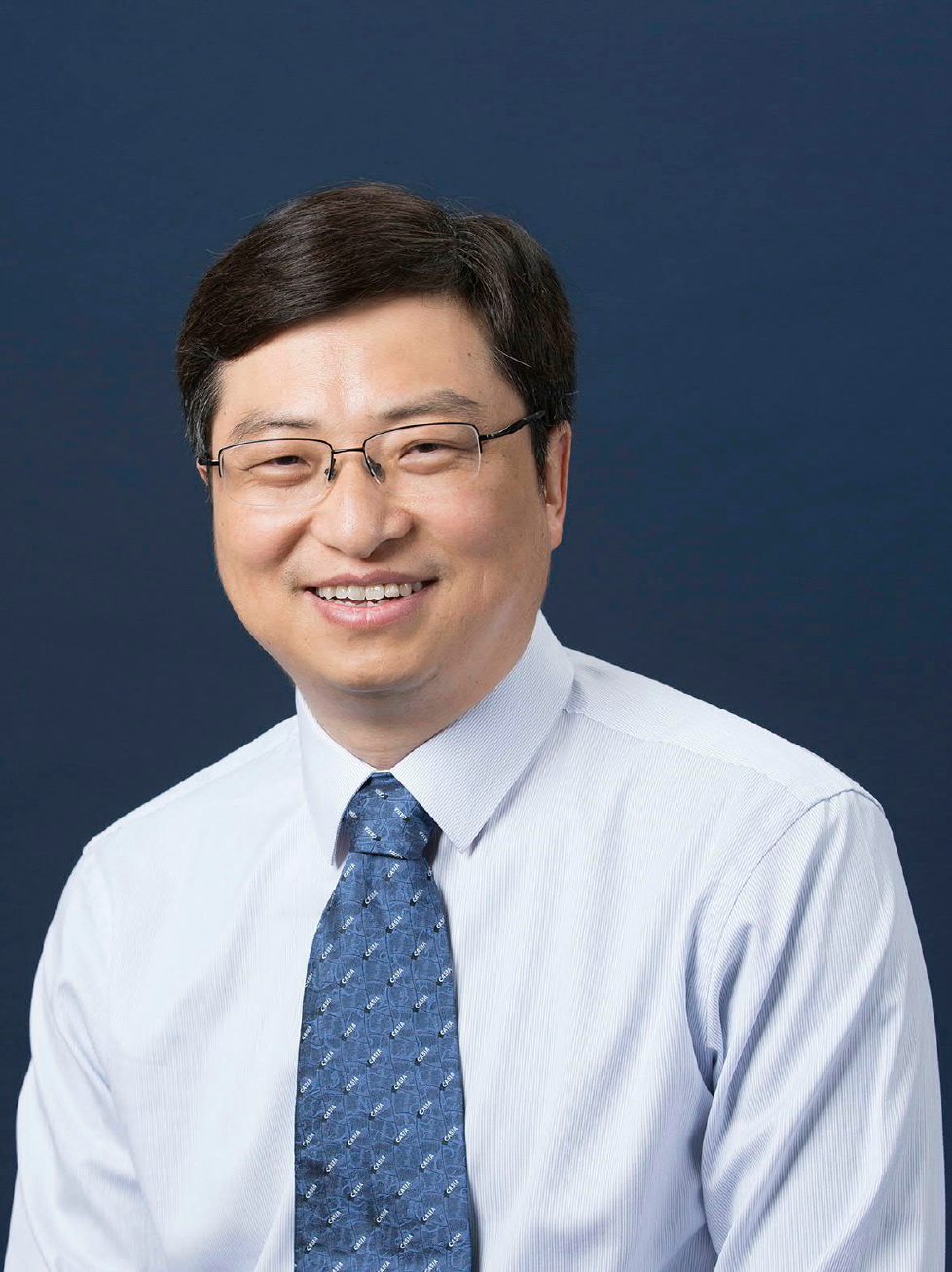}}]{Jianhua Tao}
	(Senior Member, IEEE) received the M.S. degree from Nanjing University, Nanjing, China, in 1996, and the Ph.D. degree from Tsinghua University, Beijing, China, in 2001. He is currently a Professor with NLPR, Institute of Automation, Chinese Academy of Sciences, Beijing, China. He has authored or coauthored more than 300 papers on major journals and proceedings including the IEEE TASLP, IEEE TAFFC, IEEE TIP, IEEE TSMCB, Information Fusion, etc. His current research interests include speech recognition and synthesis, affective computing, and pattern recognition. He is the Board Member of ISCA, the chairperson of ISCA SIG-CSLP, the Chair or Program Committee Member for several major conferences, including Interspeech, ICPR, ACII, ICMI, ISCSLP, etc. He was the subject editor for the Speech Communication, and is an Associate Editor for Journal on Multimodal User Interface and International Journal on Synthetic Emotions. He was the recipient of several awards from the important conferences, including Interspeech, NCMMSC, etc.
\end{IEEEbiography}
\begin{IEEEbiography}[{\includegraphics[width=1.1in,height=1.25in,clip,keepaspectratio]{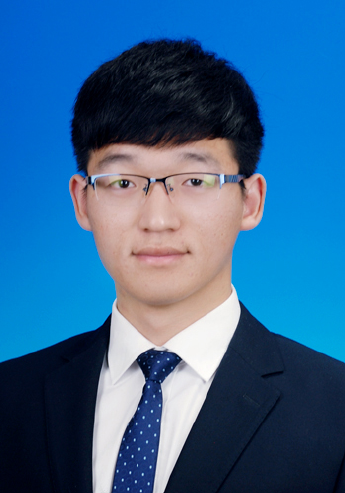}}]{Zhengkun Tian}
	received the B.S. degree from Tianjin University, Tianjin, China, in 2017. He is currently working toward the Ph.D. degree with the University of Chinese Academy of Sciences, Beijing, China. His current research interests include speech recognition, speaker verification and identification.
\end{IEEEbiography}
\begin{IEEEbiography}[{\includegraphics[width=1.1in,height=1.25in,clip,keepaspectratio]{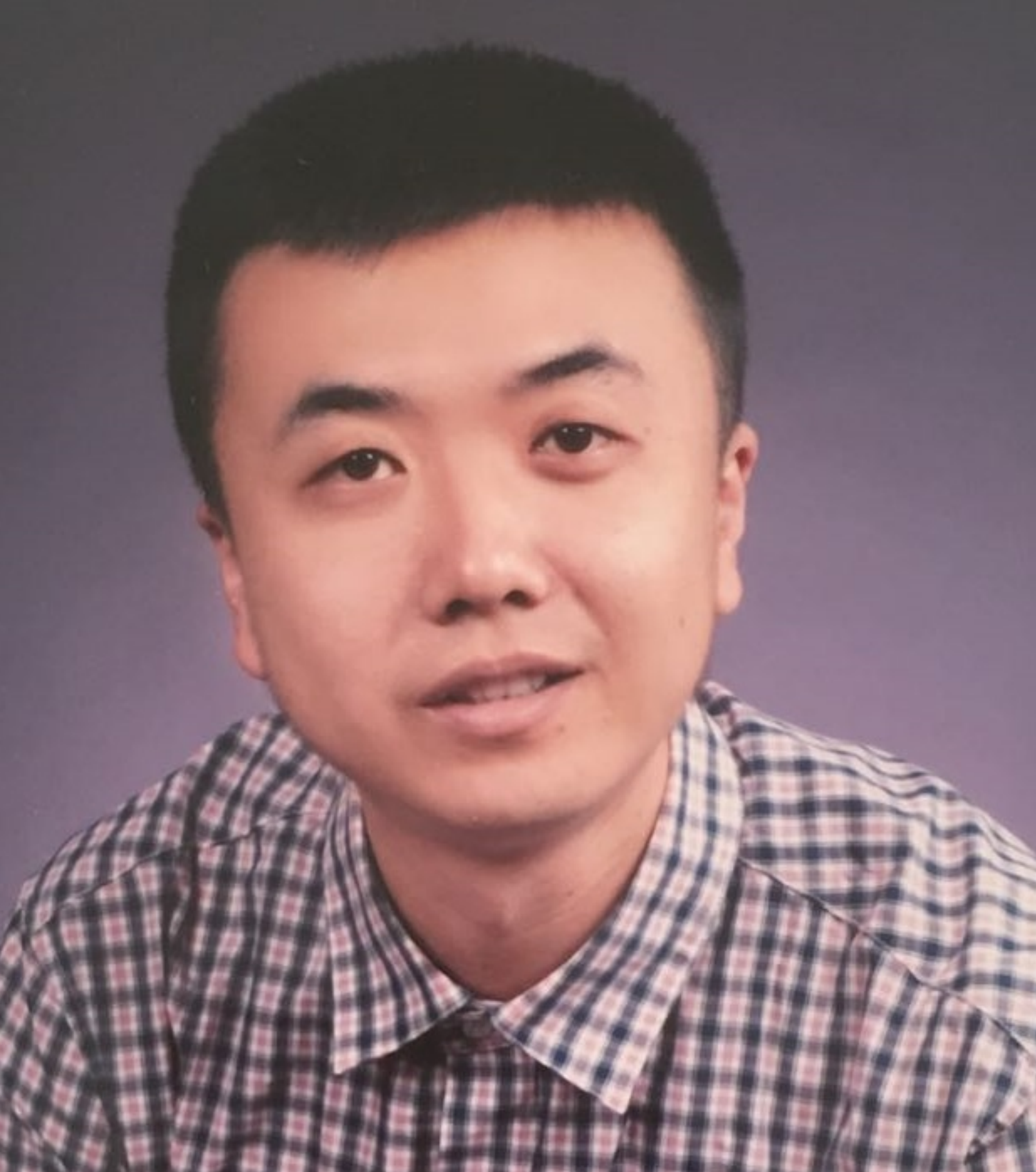}}]{Bin Liu}
	received his the B.S. degree and the M.S. degree from Beijing institute of technology (BIT), Beijing, in 2007 and 2009 respectively. He received Ph.D. degree from the National Laboratory of Pattern Recognition, Institute of Automation, Chinese Academy of Sciences, Beijing, in 2015. He is currently an Associate Professor in the National Laboratory of Pattern Recognition, Institute of Automation, Chinese Academy of Sciences, Beijing. His current research interests include affective computing and audio signal processing.
\end{IEEEbiography}
\begin{IEEEbiography}[{\includegraphics[width=1.1in,height=1.25in,clip,keepaspectratio]{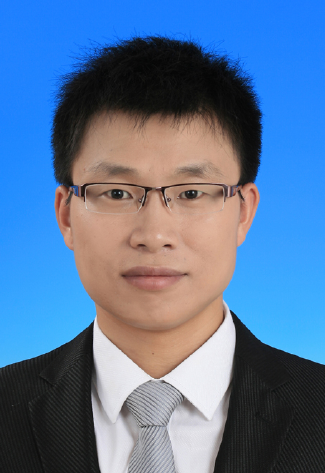}}]{Zhengqi Wen}
	received the B.S. degree from the University of Science and Technology of China, Hefei, China, in 2008, and the Ph.D. degree from the Chinese Academy of Sciences, Beijing, China, in 2013. He is currently an Associate Professor with the National Laboratory of Pattern Recognition, Institute of Automation, Chinese Academy of Sciences. His current research interests include speech processing, speech recognition, and speech synthesis.
\end{IEEEbiography}

\end{document}